\documentclass[twocolumn,showpacs,preprintnumbers,amsmath,amssymb, prb, reprint]{revtex4-1}

\usepackage{stmaryrd}
\usepackage{txfonts}
\usepackage{amssymb}
\usepackage{mathrsfs}
\usepackage{graphicx}% Include figure files
\usepackage{dcolumn}% Align table columns on decimal point
\usepackage{bm}% bold math
\usepackage{epsfig}
\usepackage{color}                    % For creating coloured text and background
\usepackage{hyperref}                 % For creating hyperlinks in cross references

\begin{document}
\preprint{\href{http://link.aps.org/doi/10.1103/PhysRevB.89.024415}{S.-Z. Lin, C. D. Batista and A. Saxena , Phys. Rev. B {\bf 89}, 024415 (2014).}}

\title{Internal modes of a skyrmion in the ferromagnetic state of chiral magnets}

\author{Shi-Zeng Lin}
\affiliation{Theoretical Division, Los Alamos National Laboratory, Los Alamos, New Mexico 87545, USA}

\author{Cristian D. Batista}
\affiliation{Theoretical Division, Los Alamos National Laboratory, Los Alamos, New Mexico 87545, USA}

\author{Avadh Saxena}
\affiliation{Theoretical Division, Los Alamos National Laboratory, Los Alamos, New Mexico 87545, USA}

\begin{abstract}
A spin texture called skyrmion has been recently observed in  certain chiral magnets without inversion symmetry. The observed skyrmions are extended objects with typical linear sizes of 10 nm to 100 nm that contain $10^3$ to $10^5$ spins and can be deformed in response to external perturbations. Weak deformations are  characterized by internal modes, which are localized around the skyrmion center. Knowledge of internal modes is crucial to assess the stability and rigidity of these topological textures. Here we compute the internal modes of a skyrmion in a ferromagnetic background state by numerical diagonalization of the dynamical matrix. We find several internal modes below the magnon continuum, such as the mode corresponding to the translational motion and different kinds of breathing modes. The number of internal modes is larger for lower magnetic fields. Indeed, several modes become gapless in the low field region indicating that the single skyrmion solution becomes unstable, although a skyrmion lattice remains thermodynamically stable. On the other hand,  only three internal modes exist at high fields and the skyrmion texture remains locally stable even when the ferromagnetic state becomes thermodynamically stable. We also show that the presence of out-of-plane easy-axis anisotropy  stabilizes the single skyrmion solution. Finally, we discuss the effects of damping and possible experimental observations of these internal modes.  
\end{abstract}
  %checked for Internal modes of a skyrmion in the ferromagnetic background in chiral magnets
\date{\today}
\maketitle

\section{Introduction}
\begin{figure}[t]
\psfig{figure=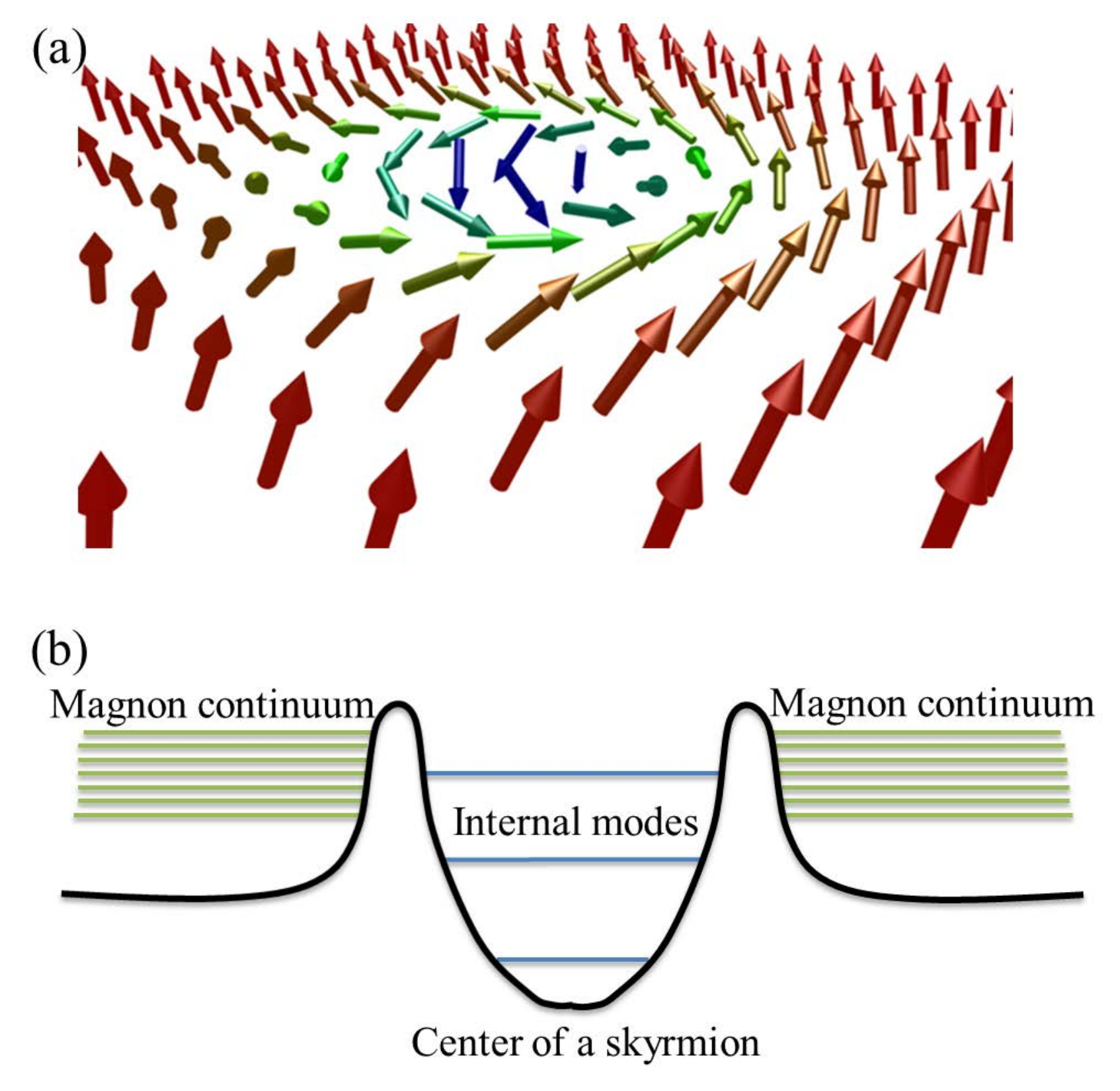,width=\columnwidth}
\caption{\label{f1}(color online) (a) Schematic view of the spin profile of a skyrmion. (b) Sketch of the effective magnon potential produced by the skyrmion. The potential is drawn along the radial direction of the skyrmion.}
\end{figure}

A skyrmion is a topological excitation which was first proposed by Skyrme a half century ago, \cite{Skyrme61} and later was observed in many condensed matter systems, such as liquid crystals, quantum Hall devices and chiral magnets. For the case of magnets, the spins wrap a sphere when moving from the center to the outer region of the skyrmion, as it is schematically shown in Fig. \ref{f1} (a). The existence of a stable skyrmion lattice in chiral magnets was predicted two decades ago.~\cite{Bogdanov89,Bogdanov94} Recently, the experimental observation of skyrmions in chiral magnets without inversion symmetry (B20 compounds), such as MnSi or $\mathrm{Fe_{0.5}Co_{0.5}Si}$, has attracted enormous attention into these emergent topological textures.~\cite{Muhlbauer2009,Yu2010a,Seki2012,Adams2012} In bulk crystals, skyrmions form a triangular lattice in a small portion of the magnetic field - temperature phase diagram. However, the skyrmion lattice phase is found to be much more stable in thin films. \cite{Yi09,Butenko2010,Yu2011,Heinze2011} 

Spontaneous skyrmion crystals have been observed  both in metals and insulators. In metals, the  skyrmion lattice can be driven by a weak current and the detrimental Joule heating can be significantly reduced.~\cite{Jonietz2010,Yu2012,Schulz2012} In insulators, the dominant dissipation mechanism is the weak Gilbert damping for spin precession. Consequently, skyrmions in insulators are very promising for applications that require low energy dissipation. Furthermore, the magnetoelectric coupling that is intrinsic to skyrmion spin textures points to a new way of controlling these magnetic topological objects with external electric fields.~\cite{Seki2012,Seki2012b} Because of these unique characteristics, skyrmions have huge potential for spintronics applications, such as information storage.~\cite{Fert2013}

Being topological textures, skyrmions cannot be created or annihilated  by continuous deformations of the spin configuration. On the other hand, the  skyrmions that have been observed in the B20 compounds are extended objects with linear size of  $10 - 100$ nm, meaning that they contain $10^3-10^5$ spins. These  skyrmions can be deformed in response to external perturbations and the actual deformation is determined by their internal modes. The number of  internal modes and their corresponding frequencies determine the stability of the skyrmion solution. In the continuum, the lowest frequency mode of the skyrmion is the gapless  Goldstone mode associated with invariance under the group of continuous translations. In recent derivations of the equation of motion  \cite{Zang11,Everschor11,Everschor12,szlin13skyrmion2,Iwasaki2013,Liu2013,Liu2013b} based on Thiele's collective coordinate approach \cite{Thiele72}, only the translational mode was taken into account by assuming that the skyrmion is rigid. Knowledge of the internal modes can test the validity  of this assumption. The internal modes also determine the scattering of  magnons by  skyrmions.

The eigenmodes of the skyrmion lattice have been calculated~\cite{Petrova2011,Mochizuki2012} and measured ~\cite{Onose2012,Okamura2013}. Three different modes with frequencies of several gigahertz have been identified. The first mode corresponds to a clockwise rotation of the skyrmion core, while the second mode corresponds to the anticlockwise rotation. The third one is a uniform breathing of all skyrmions. Excitation of these modes by strong microwave radiation leads to melting of the skyrmion lattice.~\cite{Mochizuki2012} However,  little is known about the internal modes of  a single skyrmion. The present work is devoted to filling this gap.

In the region between the skyrmion lattice and fully polarized ferromagnetic phases, the skyrmion density is low and the interaction between skyrmions  is weak. Therefore, we  can treat the skyrmions as quasi-free objects. Single skyrmions can also be created in small samples where only one skyrmion can be accommodated. \cite{szlin13skyrmion3}  Furthermore, a single skyrmion becomes a metastable solution in the ferromagnetic background of fully polarized spins that is stabilized at high magnetic fields. Consequently, the controlled manipulation of a single skyrmion in a ferromagnetic background is possible and directly relevant for applications~\cite{Romming2013}. Here we focus on this case. The magnon spectrum of the ferromagnetic state gets modified in the presence of the skyrmion. In analogy with other topological objects, such as kinks in 1D systems \cite{Currie1980,Habib1998} and vortices in 2D systems \cite{Wysin1995}, the presence of a skyrmion introduces a local potential for magnons, as it is schematically shown in Fig. \ref{f1} (b). The confining potential supports localized modes (internal modes of  the skyrmion) with energy smaller than the gap of the magnon continuum. Because the wave-length of the low-energy magnon continuum is much larger than the skyrmion size, the presence of the skyrmion has a negligible effect on the extended magnon modes.

In the present work we study the internal modes of a single skyrmion in the ferromagnetic background of a chiral magnet like the B20 compounds. We find several modes in the low magnetic field region that  become gapless at a critical magnetic field. The existence of these gapless modes renders the single skyrmion unstable, although the skyrmion crystal remains thermodynamically stable in the low field region. Only three different internal modes exist at high magnetic fields. The mode with the lowest energy is the translational mode. On a lattice, the translational mode remains gapped because the translational symmetry group is not continuous. The second mode is the nonuniform breathing in the azimuthal direction of the skyrmion. The third one is the uniform breathing mode. We show that a single skyrmion remains metastable (or locally stable) for arbitrarily  high fields. We also investigate the effect of out-of-plane easy-axis anisotropy. The skyrmion becomes more stable in the presence of easy-axis anisotropy, which could explain why the skyrmion lattice phase is more stable in thin films. The effect of damping on the internal modes is also clarified. Finally, we discuss possible experimental observations of the internal modes.

\section{Model and methods}
We consider a thin film of a chiral magnet described by the following Hamiltonian \cite{Bogdanov89,Bogdanov94,Rosler2006,Han10,Rossler2011} defined on a square lattice with  lattice constant $a$
\begin{align}\label{eq1}
\nonumber{\cal H}_s = -\frac{d}{a}\sum_{\langle i, j\rangle}\left[{J_{\rm{ex}}}{{\bf{n}}_{i}}\cdot{{\bf{n}}_{j}} + {{\bf{D}}_{ij}}\cdot({{\bf{n}}_i} \times {{\bf{n}}_{j}})\right]\\ -\frac{d}{a}\sum_i {\bf H}_a \cdot{{\bf{n}}_i}-\frac{A_z d}{2a}\sum_i n_{i,z}^2,
\end{align}
where the spins are represented by a unit vector $\mathbf{n}$ and the summation is over the nearest neighbors $\langle i,  j\rangle$. Here $\mathbf{D}_{ij}=D \hat{e}_{ij}$ is the in-plane Dzyaloshinskii-Moriya vector with $\hat{e}_{ij}$ being the unit vector from the site $i$ to the site $j$ and $A_z$ is the out-of-plane easy axis anisotropy. $J_{\rm{ex}}$ is the exchange coupling and $\mathbf{H}_a$ is the external magnetic field. We assume the system is uniform along the thickness direction so the Hamiltonian is proportional to the thickness $d$. The magnetic field is perpendicular to the film. At zero temperature ($T=0$), the skyrmion lattice phase becomes thermodynamically stable in the field region $0.2 D^2/J_{\rm{ex}}<H_a<0.8 D^2/J_{\rm{ex}}$.~\cite{Han10,Rossler2011} The magnetic spiral phase is stabilized for weak magnetic fields $H_a<0.2 D^2/J_{\rm{ex}}$, while the fully polarized ferromagnetic state becomes the ground state for $H_a>0.8 D^2/J_{\rm{ex}}$. For typical chiral magnets we have $J_{\rm{ex}}\approx 3$ meV and $D\approx 0.3$ meV. \cite{Zang11}

The spin dynamics is governed by the Landau-Lifshitz-Gilbert equation \cite{Tatara2008}
\begin{equation}\label{eq2}
{\partial _t}{\bf{n}} =- \gamma {\bf{n}} \times {\bf{H}}_{\rm{eff}} + \alpha \mathbf{n}\times {\partial _t}{\bf{n}}, 
\end{equation}
where the effective magnetic field is $\mathbf{H}_{\rm{eff}}\equiv-\delta \mathcal{H}/\delta {\bf{n}}$ and $\gamma=1/(\hbar s)$ with $s$ being the magnitude of the local spins. The Gilbert damping coefficient $\alpha$ is weak for real materials, $\alpha\ll 1$.

We find a stationary (metastable) solution of a single skyrmion in the ferromagnetic background. On a lattice, the energy minimized when the center of the skyrmion is located at a high symmetry point, i.e., one lattice site or the center of a square plaquette. We find that the stationary solution corresponds to the case in which the skyrmion is centered  at a lattice site. The unperturbed skyrmion is circular and can be parametrized by $\mathbf{n}_s=(\sin\theta_s\cos\phi_s,\ \sin\theta_s\sin\phi_s,\ \cos\theta_s)$ with $\phi_s=\mathrm{atan}[(y-y_0)/(x-x_0)]+\pi/2$. We use the relaxation method to find the solution for $\theta_s$.  In the continuum limit ($a\rightarrow 0$), $\theta_s$  is determined by
\begin{equation}\label{eq3}
\frac{\partial \theta_s }{\partial t}=-\left[\cos(2 \theta_s ) +\frac{\sin (2\theta_s)}{2 r}+\frac{\beta}{2} r \sin\theta_s-\frac{\partial \theta_s }{\partial r}-1-r\partial _r^2\theta_s \right],
\end{equation}
where $\beta=2H_a J_{\mathrm{ex}}/D^2$ and length is in units of $J_{\mathrm{ex}} a/D$. Here $\theta_s=\pi$ at the center of the skyrmion $(x_0,\ y_0)$ and decays to $\theta_s=0$ far away from the skyrmion. 

To calculate the internal modes, we add a small perturbation to the stationary solution: $\mathbf{n}=\mathbf{n}_s+\tilde{\mathbf{n}}$ with $\tilde{\mathbf{n}}\ll 1$. Because $|n|$ is conserved,  to the lowest order $\tilde{\mathbf{n}}$ is perpendicular to $\mathbf{n}_s$. We use a local coordinate approach \cite{Wysin1995} in which the local $\tilde{z}$ axis is parallel to the $\mathbf{n}_s$ direction, while $\tilde{\mathbf{n}}$ lies in the local $\tilde{x}$-$\tilde{y}$ plane. We then substitute $\mathbf{n}_s$ into Eq.~\eqref{eq2} and keep the contributions of  first order  in $\tilde{\mathbf{n}}$. Equation ~\eqref{eq2} can be expressed as a matrix equation in the frequency domain and we obtain the lowest 30 eigenfrequencies and eigenmodes by diagonalizing the matrix with  the Lanczos method. Details of this calculation are included in Appendix A.

\section{Results}
\begin{figure}[t]
\psfig{figure=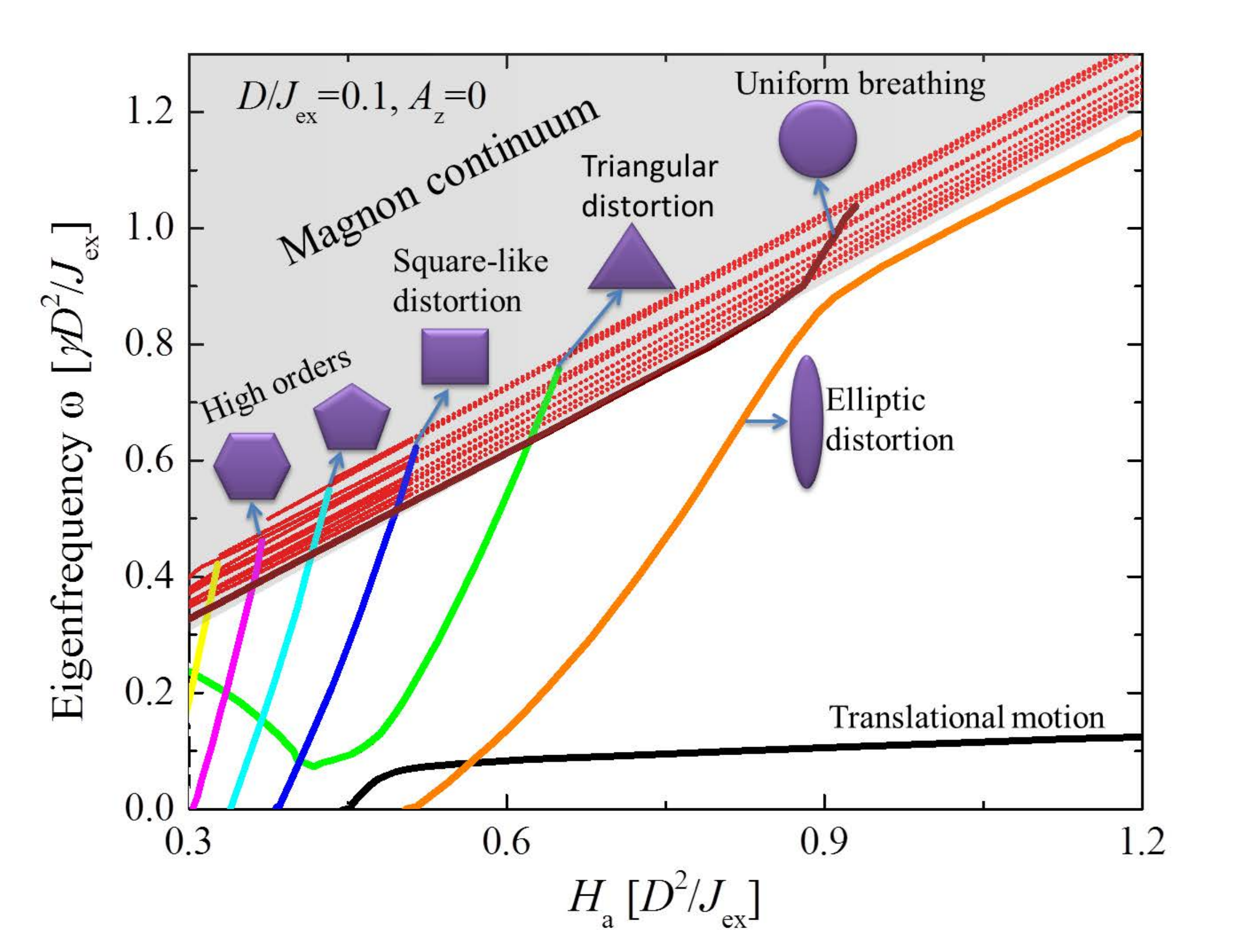,width=\columnwidth}
\caption{\label{f2}(color online) Magnetic field dependence of the eigenfrequencies of different modes in the absence of easy-axis anisotropy ($A_z=0$). The modes are assigned to the $l$-th order breathing mode and the translational mode. We show only several eigenfrequencies of the magnon continuum (shaded region)  because  only the lowest $30$ eigenmodes have been obtained with  the Lanczos method.}
\end{figure}

\begin{figure}[t]
\psfig{figure=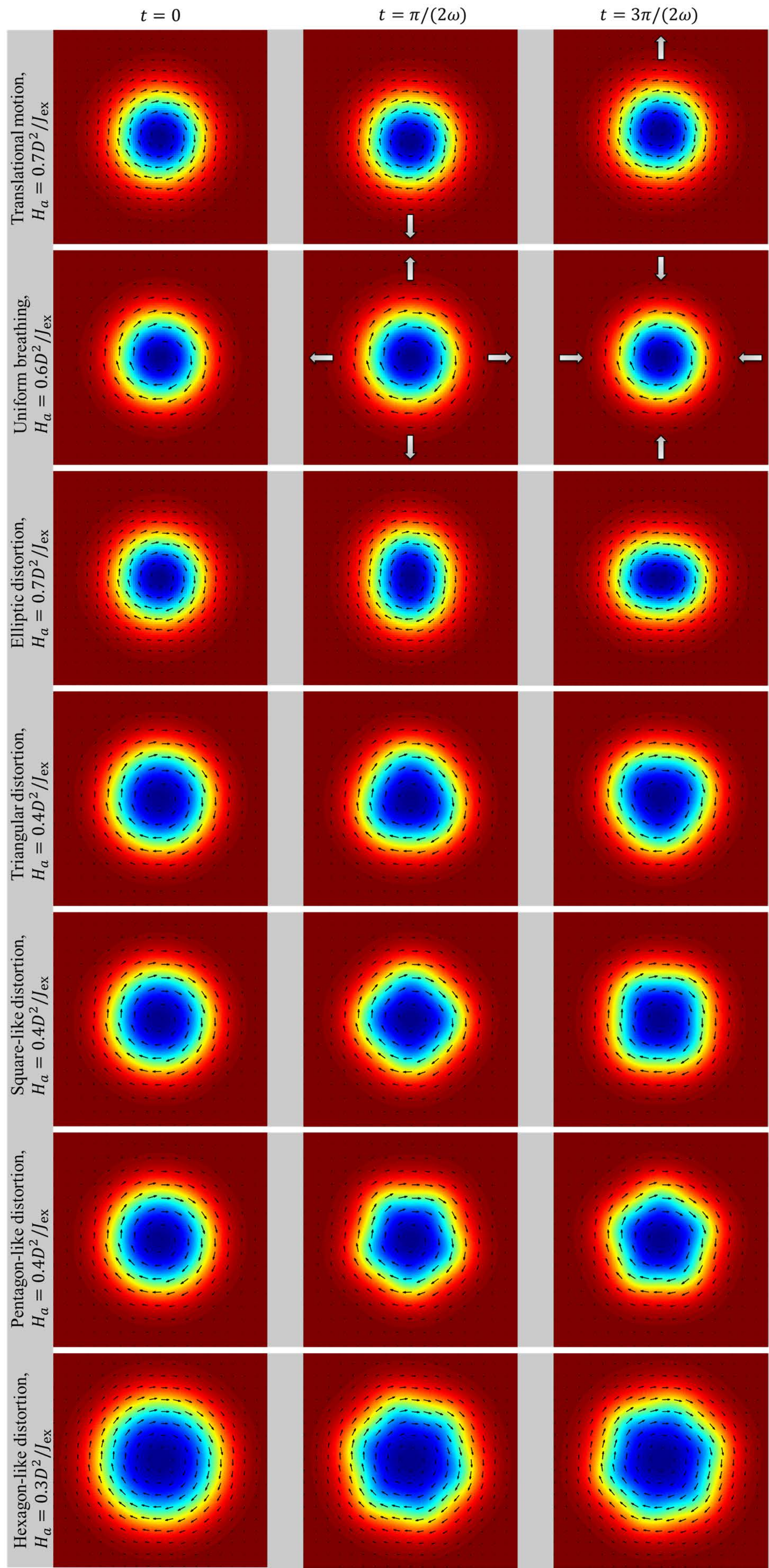,width=0.92\columnwidth}
\caption{\label{f3}(color online) Deformation of the skyrmion associated with different internal modes. The spin configuration is obtained from the eigenvector $\tilde{\mathbf{n}}_v$ by using $\mathbf{n}=\mathbf{n}_s+F \tilde{\mathbf{n}}_v\sin(\omega t)$ with $F=10$. The spin configuration is plotted at $t=0$ (left column), $t=\pi/(2\omega)$ (middle column) and $t=3\pi/(2\omega)$ (right column). The white arrows in the first row denote the direction of the translational motion and in the second row represent the direction of the uniform breathing. The vectors in the plots denote the $n_x$ and $n_y$ components, and the contour plot  denotes the $n_z$ component with red representing $n_z=1$ and blue representing $n_z=-1$. Here $D/J_{\mathrm{ex}}=0.1$ and $A_z=0$.}
\end{figure}

The magnetic field dependence of the low eigenfrequencies is shown in Fig.~\ref{f2} for $A_z=0$ and  $D/J_{\mathrm{ex}}=0.1$. To trace the continuous evolution of the modes as a function of $H_a$, the eigenfrequencies are calculated at field intervals as small as  $\delta H_a=0.00125$. The applied magnetic field opens a gap, $\omega_g=\gamma H_a$, in the magnon spectrum.   If we adopt the boundary condition  $\mathbf{\tilde{n}}({\bf r})=0$ for ${\bf r} $ on the boundary, the lowest wavenumber is $k=\pi/L$ (where $L$ is the linear size of the lattice). \footnote{ The minimal wavenumber is $k=0$ for open boundary conditions ( $\partial_\mathbf{r}\mathbf{n}=0$ at the boundary) in the continuum limit $a\rightarrow 0$ and the magnon continuum begins at $\omega_g$. Here $\mathbf{n}$ is a unit vector normal to the boundary.} Correspondingly, the lowest frequency of the magnon continuum is slightly higher than $\omega_g$. To assign the internal modes, we calculate the profile of the spin configuration $\mathbf{n}=\mathbf{n}_s+F \tilde{\mathbf{n}}_v\sin(\omega t)$ where $\tilde{\mathbf{n}}_v$ is the eigenvector, $F$ is an arbitrary amplitude and $\omega$ is the eigenfrequency. Snapshots for $t=0$, $t=\pi/(2\omega)$ and $t=3\pi/(2\omega)$  are shown in Fig. \ref{f3} for several typical modes. The translational mode exists in the whole magnetic field region.  The skyrmion gets deformed into $l$-th order polygons. For instance,  for $l=2$ it deforms into an ellipse with its major axis along the $y$ direction in the first half period and along the $x$ direction in the next half period. We will refer to these modes as  $l$-th order breathing modes with the uniform breathing mode corresponding to $l=0$.

\begin{figure}[t]
\psfig{figure=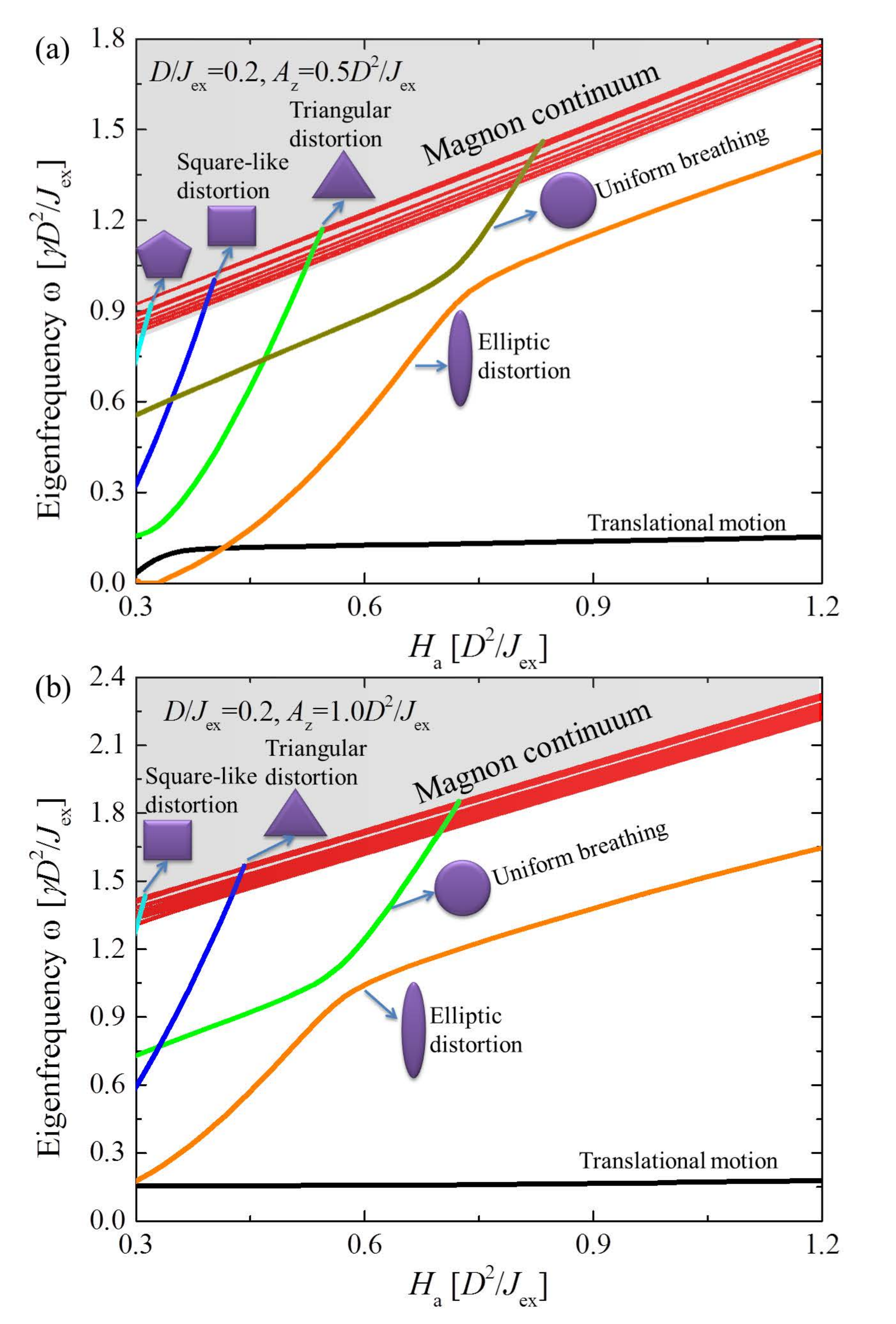,width=0.9\columnwidth}
\caption{\label{f4}(color online) Same as Fig. \ref{f2} but with $A_z=0.5D^2/J_{\mathrm{ex}}$ in (a) and $A_z=1.0D^2/J_{\mathrm{ex}}$ in (b). The eigenfrequencies for $ D/J_{\mathrm{ex}}=0.1$ and $0.2$ are almost identical in normalized units.}
\end{figure}

The  skyrmion size depends on the magnetic field value. Because the skyrmion is bigger for weak fields,  it supports more internal modes. As shown in Fig.~\ref{f2},  some levels cross as a function of field indicating that these modes become degenerate at the crossing magnetic field values. Each $l$-th order breathing mode becomes gapless at a specific magnetic field $H_l$, indicating an instability towards the deformation associated with such mode. Below the threshold magnetic field $H_l$, the eigenfrequency of the $l$-th order breathing mode becomes negative. The effective damping, which is proportional to the gap, also becomes negative [see Eq.~\eqref{eq4} below]. Thus, the amplitude of the mode increases exponentially with time  indicating that the skyrmion configuration is no longer stable. It is important to note that the skyrmion solution becomes unstable at a single critical field for which the first mode  becomes gapless. Because the first mode that becomes soft is  the elliptic ($l=2$) mode, the skyrmion solution is locally stable only for $H_a > H_2$. The softening of $l=2$ mode indicates that the skyrmion becomes unstable against elliptical distortions. Therefore, the spectrum of eigenfrequencies shown in Fig.~\ref{f2}
is not  is not physically accessible for $H_a<H_2$ unless the skyrmion is confined in a geometry that makes it stable against elliptical distortions. The same reasoning applies to the successive modes that become soft at fields lower than $H_2$ [see Fig.~\ref{f2}].
To verify the stability of the skyrmion solution at low fields, we performed additional simulations with one skyrmion at the center of the ferromagnetic state as the initial state.  By solving Eqs.~\eqref{eq1} and \eqref{eq2} numerically, we found that the skyrmion solution becomes unstable relative to weak perturbations for fields $H_a\lesssim0.55D^2/J_{\mathrm{ex}}$ in agreement with our estimation of 
$H_2$ based on the softening of the $l=2$ mode. 

Three different internal modes appear below the magnon continuum at high fields $H_a\gtrsim 0.55D^2/J_{\mathrm{ex}}$.~\footnote{Without anisotropy ($A_z=0$) the uniform breathing mode is slightly above the magnon gap $\omega_g$ and its frequency increases rapidly for higher fields. With anisotropy ($A_z>0$), the uniform breathing mode appears below the magnon gap and it extends into the magnon continuum for higher fields (see Fig. \ref{f4}).} The lowest frequency mode corresponds
to a uniform translation  of the skyrmion. This translational mode has a non-zero frequency because of the intrinsic pinning caused by the discrete lattice. The frequency depends on the ratio of the skyrmion size to the lattice constant and it becomes bigger for higher fields  because the skyrmion becomes smaller. We also computed the dependence of this frequency on $D$ and verified that it is proportional to $D^2$ because the skyrmion size shrinks for a larger $D$.  The mode with the second lowest frequency is the $l=2$ breathing mode. Its frequency increases rapidly with the magnetic field and finally approaches the magnon continuum. The  frequency of the uniform breathing mode ($l=0$)  is above the magnon gap for high fields implying that this mode gets buried inside the magnon continuum. 

We next discuss the effect of the easy-axis anisotropy $A_z$ on the internal modes. The results for $A_z=0.5D^2/J_{\mathrm{ex}}$ and $A_z=1.0D^2/J_{\mathrm{ex}}$ are depicted in Figs.~\ref{f4} (a) and (b). The presence of anisotropy lifts the magnon gap to the frequency $\omega_g=\gamma({H_a+A_z})$ and the uniform breathing mode is now below and well separated from the magnon continuum for low fields. The number of internal modes that lie inside the gap is reduced at low fields. The skyrmion size shrinks for a large anisotropy and the effective potential for magnons becomes narrower, which lifts the eigenfrequencies of the internal modes towards the magnon continuum.  The skyrmion becomes stable over  a wider region of magnetic fields. The effect of  out-of-plane easy-axis anisotropy is similar to the effect of the external magnetic field, which  stabilizes the skyrmion texture.

\begin{figure}[t]
\psfig{figure=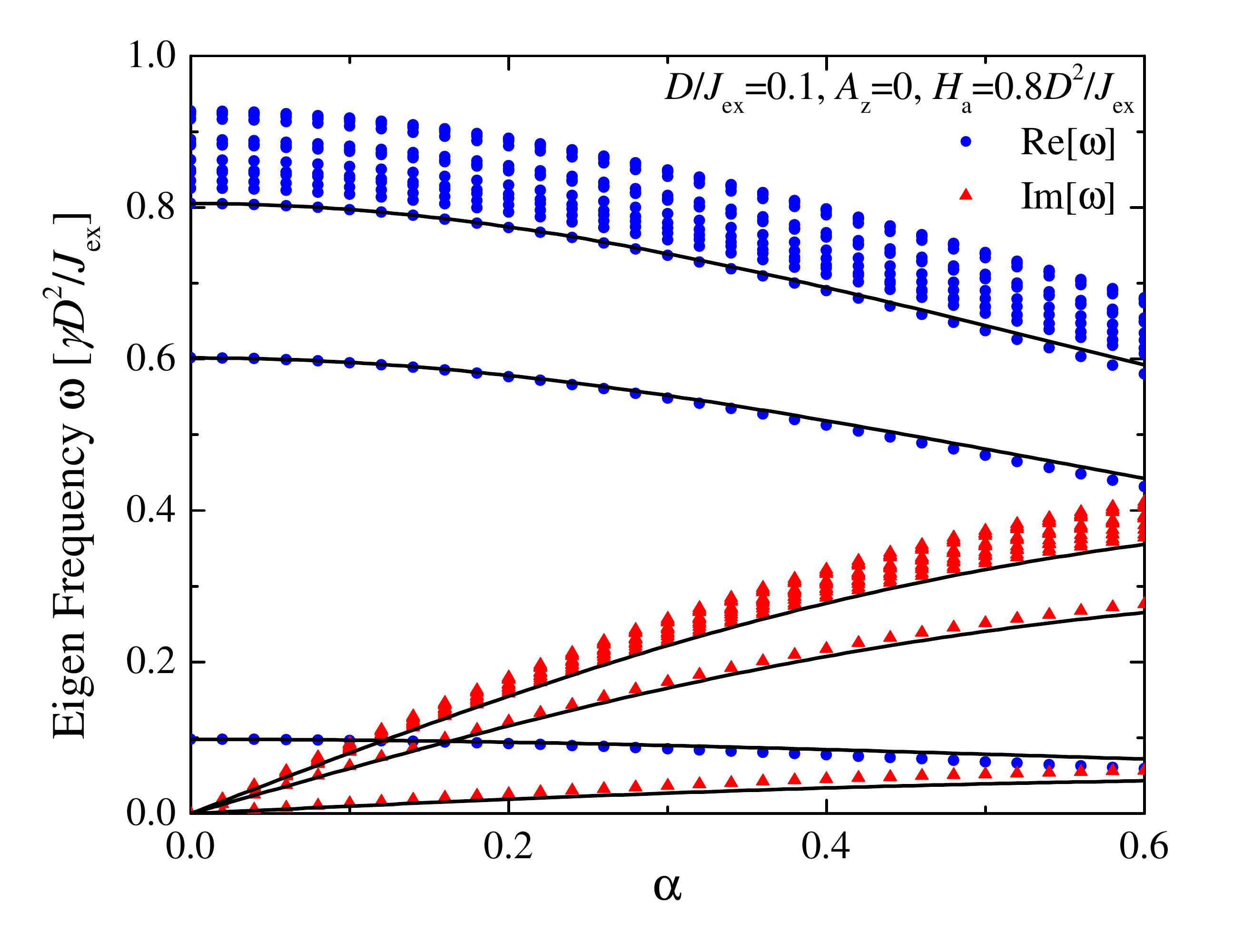,width=\columnwidth}
\caption{\label{f5}(color online) Dependence of the eigenfrequencies $\omega$ on the damping coefficient $\alpha$. Symbols are from numerical diagonalization and lines are given by Eq. \eqref{eq4}. For clarity, not all lines are shown.}
\end{figure}

Finally we study the effect of the damping $\alpha$ on the internal modes. With damping, the eigenfrequencies  acquire an imaginary part meaning that the lifetime of  the modes becomes finite: $\tau=2\pi/\mathrm{Im}[\omega]$. The dependence of the real, $\mathrm{Re}[\omega]$, and imaginary part, $\mathrm{Im}[\omega]$, of the eigenfrequencies on $\alpha$ is shown in Fig.~\ref{f5}. For $\alpha\ll 1$, this dependence is  well described by 
\begin{equation}\label{eq4} 
\omega(\alpha)=\frac{1+\alpha i}{1+\alpha^2}\omega(\alpha=0),
\end{equation}
which is the dependence that is obtained for the ferromagnetic state. 

\section{Discussion}
In the low field region where the skyrmion lattice is thermodynamically stable, the single skyrmion solution in a ferromagnetic background is, however, unstable. One possible reason is that in the skyrmion lattice, the mutual repulsion between skyrmions suppresses the internal modes associated with skyrmion distortion and stabilizes the skyrmion lattice. A single skyrmion is only stable at relatively high fields. The single skyrmion solution becomes  a metastable state at higher fields, when the ferromagnetic state becomes thermodynamically stable. The creation of a single skyrmion  requires to overcome a large energy barrier deep inside the ferromagnetic phase. Thus single skyrmion manipulation becomes easier near the boundary between the ferromagnetic and skyrmion crystal phases.~\cite{Romming2013} It has been demonstrated that an out-of-plane easy axis is helpful to stabilize the skyrmion state. Because this anisotropy is inversely proportional to the film thickness \cite{Johnson1996}, $A_z \propto 1/d$, skyrmions are more stable in thin films. \cite{Yi09,Butenko2010,Yu2011,Heinze2011} 

The equation of motion for a rigid skyrmion has been derived by several authors based on Thiele's collective coordinate approach, where only the translational mode is taken into account. According to Fig. \ref{f2}, the translational mode has the lowest energy for a stable skyrmion and it is well separated from the other modes. Thus, the equation of motion for a rigid skyrmion is quite accurate, as it was confirmed by different numerical simulations. Higher order corrections, which lead to the generation of a finite skyrmion mass caused by deformations, are expected to be small.

The mass of a magnetic bubble domain, which has the same skyrmion topological number as the spin configuration considered here, has been calculated in Ref. \onlinecite{Makhfudz2012}. The result is that a magnetic bubble domain has a sizeable mass. We would like to point out that the magnetic bubble considered in Ref.~\onlinecite{Makhfudz2012} is different from the spin texture that we have considered here. In Ref. \onlinecite{Makhfudz2012}, the spins point down inside the bubble and up outside the bubble. In our case, the spins  gradually change from up to down  and they simultaneously rotate along a  direction determined by the Dzyaloshinskii-Moriya interaction when moving away from the center of the skyrmion. Because  the  mass arises from distortions induced by the skyrmion motion, the masses for these two different spin configurations can differ drastically because their internal modes can have very different frequencies. By comparing our results with those in Ref. \onlinecite{Makhfudz2012}, we can infer that skyrmions of chiral magnets are much more rigid  than  bubble domains. 

In addition, the authors of Ref.~\onlinecite{Makhfudz2012} have \emph{only} considered distortions  of the domain wall, which can be described by two waves moving around the wall in the opposite directions. To certain extent, the  skyrmion distortion shown in Fig.~\ref{f3}  can also be described by two waves moving around the core  in opposite directions. However,  the distortion extends over the whole region of the skyrmion as it is shown in Fig.~\ref{f3}. Moreover, we have shown explicitly that the elliptic distortion ($l=2$) has the lowest eigenfrequency implying that this mode provides the dominant contribution to the skyrmion mass in chiral magnets.

We next discuss possible experimental observations of the internal modes. The internal modes have frequencies of several gigahertz. Thus, they can be  excited  and observed with microwave absorption measurements. Because the wavelength of microwaves in the gigahertz region is much larger than the linear skyrmion size, only the uniform breathing mode can be significantly excited. The other modes with $l>0$ can be measured with a local probe, such as spin polarized scanning tunneling microscope.

\section{Conclusion}
To summarize, we have calculated the internal modes of a single skyrmion in the ferromagnetic state of chiral magnets. A translational mode and different types of  breathing modes appear below the magnon continuum. Several modes become gapless in the low field region and the single skyrmion becomes unstable in spite of the fact that the skyrmion lattice is thermodynamically stable in this field region. The single skyrmion (excited) state is locally stable for high fields including the field region for which the fully saturated ferromagnetic state is thermodynamically stable. Out-of-plane easy-axis anisotropy increases the local stability of the single skyrmion state.  The translational mode is gapped for spins  on a lattice and the gap decreases rapidly as a function of increasing ratio between the linear  skyrmion size and  the lattice constant. We find that  the translational mode has the lowest energy and it is well separated from the other modes. This result  justifies the rigid skyrmion approximation in derivation of the particle-like equation of motion.  We have also discussed the effects of damping on the eigenfrequencies and  possible experimental probes for measuring  the internal modes of a single skyrmion.

\section{Acknowledgments}
We thank Lev N. Bulaevskii, Charles Reichhardt, Yasuyuki Kato, Yoshitomo Kamiya and Oleg Tchernyshyov for useful discussions. This work was supported by the U.S. Department of Energy.

\appendix

\begin{figure}[b]
\psfig{figure=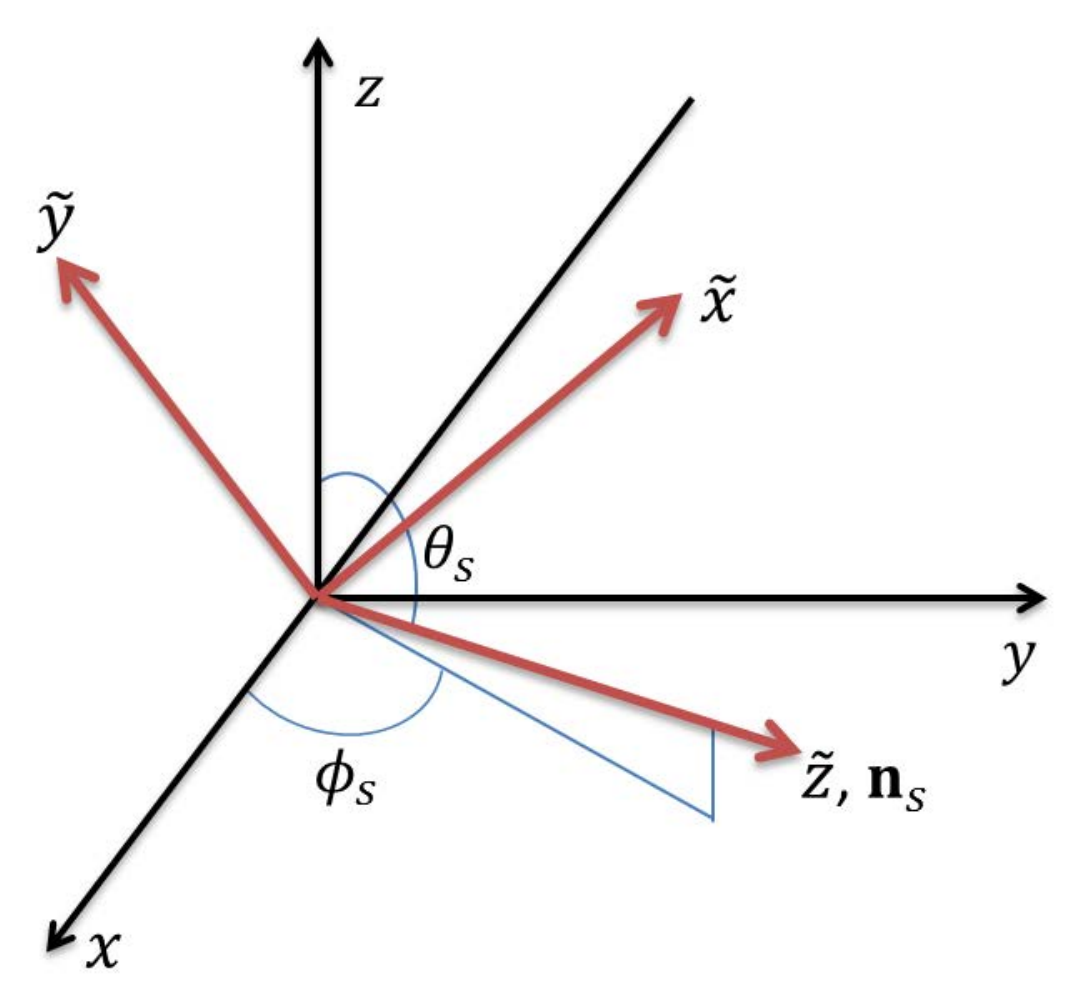,width=\columnwidth}
\caption{\label{f6}(color online) Definition of the local coordinate system. The local $\tilde{z}$ axis is along the spin direction of the unperturbed skyrmion. The local $\tilde{x}$ axis is in the $x$-$y$ plane of the original coordinates.}
\end{figure}

\section{Equation of motion for perturbations $\tilde{n}$}

The internal modes of a skyrmion cannot be obtained by expanding in variations of the azimuthal and polar angles, $\theta=\theta_s+\tilde{\theta}$ and $\phi=\phi_s+\tilde{\phi}$,  because the expansion is ill defined at the poles $\theta_s=0$ or $\theta_s=\pi$. The resulting equation of motion includes a term $\partial_t\tilde{\phi}\sin\theta_s$ and $\tilde{\phi}$  is not necessarily small when the spin is close to any of the two poles.  This singular behavior arises from the topological nature of SU(2) [SU(2) is a closed manifold] and  it cannot be avoided  by using fixed Cartesian coordinates, because the spins of a skyrmion point in all possible directions.
To overcome this problem, we use a local coordinate approach,~\cite{Wysin1995} in which the local $\tilde{z}$ axis is parallel to the local spin direction of the unperturbed skyrmion. The local $\tilde{x}$ axis is in the $x$-$y$ plane of the original spin coordinates, as it shown in Fig.~\ref{f6}. To linear order, the perturbation  $\tilde{\mathbf{n}}$ lies in the $\tilde{x}$-$\tilde{y}$ plane. The  original, $\mathbf{n}$, and the local spin coordinates, $\tilde{\mathbf{n}}$, are connected by the rotation
\begin{widetext}
\begin{equation}\label{eqa1}
\left( {\begin{array}{*{20}{c}}
{{n_x}}\\
{{n_y}}\\
{{n_z}}
\end{array}} \right) = \left[ {\begin{array}{*{20}{c}}
{ - \sin {\phi _s}}&{ - \cos {\phi _s}\cos {\theta _s}}&{\sin {\theta _s}\cos {\phi _s}}\\
{\cos {\phi _s}}&{ - \sin {\phi _s}\cos {\theta _s}}&{\sin {\theta _s}\sin {\phi _s}}\\
0&{\sin {\theta _s}}&{\cos {\theta _s}}
\end{array}} \right]\left( {\begin{array}{*{20}{c}}
{{{\tilde n}_x}}\\
{{{\tilde n}_y}}\\
{{{\tilde n}_z}}
\end{array}} \right),
\end{equation}
where $\phi_s$ and $\theta_s$ are the phases of a stationary skyrmion. $\tilde{n}_z=1$ to linear order in  $\tilde{n}_x,\ \tilde{n}_y\ll 1$. For convenience, we denote $\theta_s(i', j')$ at the site $(i',\ j')$ in the square lattice as $\theta_{s, \mathbf{ i}}$, and similarly for $\phi_{s, \mathbf{ i}}$. We also introduce
\begin{equation}\label{eqa2}
s_{\mathbf{ i}}=\sin\theta_{s, \mathbf{ i}},\ \ \ \ \ \  c_{\mathbf{ i}}=\cos\theta_{s, \mathbf{ i}}.
\end{equation}

Substituting Eq. \eqref{eqa1} into Eq. \eqref{eq1}, we obtain the following zeroth order contribution to the Hamiltonian per unit of thickness
\begin{align}
\nonumber\mathcal{H}^{(0)}=\sum_{\mathbf{i}}\sum_{\mathbf{j}=\mathbf{i}+\hat{e}_x,\mathbf{i}+\hat{e}_y }J_{\mathrm{ex}}\left[-s_{\mathbf{i}}s_{\mathbf{j}}\cos(\phi_{\mathbf{i}}-\phi_{\mathbf{j}})-c_{\mathbf{i}}c_{\mathbf{j}}\right]-\sum_{\mathbf{i}}\left[H_a c_{\mathbf{i}}+\frac{A_z}{2}c_{\mathbf{i}}^2\right]\\
+\sum_{\mathbf{i}, \mathbf{j}=\mathbf{i}+\hat{e}_x}D\left[c_{\mathbf{i}} s_{\mathbf{j}} \sin\phi_{\mathbf{j}}-s_{\mathbf{i}} c_{\mathbf{j}} \sin\phi_{\mathbf{i}}\right]+\sum_{\mathbf{i}, \mathbf{j}=\mathbf{i}+\hat{e}_y}D\left[-c_{\mathbf{i}} s_{\mathbf{j}} \cos\phi_{\mathbf{j}}+s_{\mathbf{i}} c_{\mathbf{j}} \cos\phi_{\mathbf{i}}\right].
\end{align}
The second order contribution, $\mathcal{H}^{(2)}=\mathcal{H}_{\mathrm{ex}}^{(2)}+\mathcal{H}_{\mathrm{DM,x}}^{(2)}+\mathcal{H}_{\mathrm{DM,y}}^{(2)}+\mathcal{H}_{\mathrm{A}}^{(2)}$, is given by 
\begin{align}
\mathcal{H}_{\mathrm{ex}}^{(2)}=-J_{\mathrm{ex}}\sum_{\mathbf{i}}\sum_{\mathbf{j}=\mathbf{i}+\hat{e}_x,\mathbf{i}+\hat{e}_y }\left[{\tilde{n}_{y,\mathbf{i}}}\left({\tilde{n}_{y,\mathbf{j}}}(c_\mathbf{i} c_\mathbf{j}\cos(\phi_\mathbf{i} -\phi_\mathbf{j}) + s_\mathbf{i} s_\mathbf{j}) - c_\mathbf{i} {\tilde{n}_{x,\mathbf{j}}}\sin({\phi _\mathbf{i}} - {\phi _\mathbf{j}})\right) + \tilde{n}_{x,\mathbf{i}}\left(\cos(\phi _\mathbf{i} - \phi _\mathbf{j})\tilde{n}_{x,\mathbf{j}} + c_\mathbf{j} \tilde{n}_{y,\mathbf{j}} \sin(\phi_\mathbf{i} - \phi_\mathbf{j})\right)\right],
\end{align}
\begin{align}
\mathcal{H}_{\mathrm{DM,x}}^{(2)}=-\sum_{\mathbf{i}, \mathbf{j}=\mathbf{i}+\hat{e}_x}D\left[ - \cos\phi_\mathbf{j} s_\mathbf{i} \tilde{n}_{y,\mathbf{i}} \tilde{n}_{x,\mathbf{j}} + \tilde{n}_{y,\mathbf{j}}\left(\cos{\phi_\mathbf{i}}s_\mathbf{j} \tilde{n}_{x,\mathbf{i}} + \tilde{n}_{y,\mathbf{i}}( - c_\mathbf{i} s_\mathbf{j} \sin{\phi_\mathbf{i}} + c_\mathbf{j} s_\mathbf{i} \sin{\phi _\mathbf{j}})\right)\right],
\end{align}
\begin{align}
\mathcal{H}_{\mathrm{DM,y}}^{(2)}=-\sum_{\mathbf{i}, \mathbf{j}=\mathbf{i}+\hat{e}_y}D\left[\sin\phi_\mathbf{i} s_\mathbf{j} \tilde{n}_{y,\mathbf{j}} \tilde{n}_{x,\mathbf{i}} - \tilde{n}_{y,\mathbf{i}}\left(\sin{\phi_\mathbf{j}}s_\mathbf{i} \tilde{n}_{x,\mathbf{j}} + \tilde{n}_{y,\mathbf{j}}( - c_\mathbf{i} s_\mathbf{j} \cos{\phi_\mathbf{i}} + c_\mathbf{j} s_\mathbf{i} \cos{\phi _\mathbf{j}})\right)\right],
\end{align}
\begin{align}
\mathcal{H}_{\mathrm{A}}^{(2)}=-\frac{A_z}{2}\sum_{\mathbf{i}} \left(s_\mathbf{i}\tilde{n}_{y,\mathbf{i}}\right)^2.
\end{align}
\end{widetext}
Here $\hat{e}_x$ and $\hat{e}_y$ are  unit vectors along the $x$ and $y$ directions respectively. The first order contribution does not enter into the equation of motion for $\tilde{n}_x$ and $\tilde{n}_y$. The effective field is then  given by $\tilde{H}_{\mathrm{eff}, x}=-\delta \mathcal{H}^{(2)}/\delta\tilde{n}_x$, $\tilde{H}_{\mathrm{eff}, y}=-\delta \mathcal{H}^{(2)}/\delta\tilde{n}_y$ and $H_{\mathrm{eff}, z}=-\mathcal{H}^{(0)}$. The equation for $\tilde{\mathbf{n}}$ in the frequency domain $\tilde{\mathbf{n}}(t)\sim \tilde{\mathbf{n}}(\omega)\exp(i\omega t)$ is
\begin{equation}\label{eqa8}
\frac{i\omega}{\gamma} (\tilde{n}_{x}+\alpha\tilde{n}_{y})=\tilde{H}_{\mathrm{eff}, y}-\tilde{n}_{y}H_{\mathrm{eff}, z} ,
\end{equation}
\begin{equation}\label{eqa9}
\frac{i\omega}{\gamma} (\tilde{n}_{y}-\alpha\tilde{n}_{x})=-\tilde{H}_{\mathrm{eff}, x}+\tilde{n}_{x}H_{\mathrm{eff}, z} .
\end{equation}
We now introduce the vector $\mathbf{v}_x=(\tilde{n}_{x, 1},\ \tilde{n}_{x, 2},\ \tilde{n}_{x, 3}, \ \cdots)^T$ and similarly for  $\mathbf{v}_y$ in order to rewrite Eqs~\eqref{eqa8} and \eqref{eqa9} in a matrix form
\begin{equation}
\frac{{{{i}}\omega }}{\gamma }\left(1 + {\alpha ^2}\right)\left( {\begin{array}{*{20}{c}}
{{\mathbf{v}_x}}\\
{{\mathbf{v}_y}}
\end{array}} \right) = \mathbf{M}\left( {\begin{array}{*{20}{c}}
{{\mathbf{v}_x}}\\
{{\mathbf{v}_y}}
\end{array}} \right).
\end{equation}
In the absence of damping ($\alpha=0$), this matrix is antisymmetric, $\mathbf{M}=-\mathbf{M}^T$,  because $\omega$ is real. We obtain the thirty lowest eigenfrequencies, $\omega$, and corresponding eigenvectors of $\mathbf{M}$ by using the Lanczos method. The lattice size that we use for numerical diagonalization is  much larger than the linear skyrmion size to eliminate spurious size effects. However, the extended modes still  depend on the system size. Typically we use 400 lattice constants along the $x$ and $y$ directions. The corresponding size of the matrix $\mathbf{M}$ is $320000\times320000$. 

%\bibliography{reference}

%merlin.mbs apsrev4-1.bst 2010-07-25 4.21a (PWD, AO, DPC) hacked
%Control: key (0)
%Control: author (8) initials jnrlst
%Control: editor formatted (1) identically to author
%Control: production of article title (-1) disabled
%Control: page (0) single
%Control: year (1) truncated
%Control: production of eprint (0) enabled
%

\end{document}